\documentstyle[aps,prb,epsf]{revtex}
\tighten
\begin{document}
\draft

\title{Directional Emission from a Microdisk Resonator
with a Linear Defect }
\author{V. M. Apalkov and M. E. Raikh}
\address{Department of Physics, University of Utah, Salt Lake City,
Utah 84112}

\maketitle
\begin{abstract}
Microdisk resonator with a linear defect at some distance away
from the circumference is studied theoretically.
We demonstrate that the presence of the defect leads to ({\em i})
 enhancement of the output efficiency, and  ({\em ii})
 directionality of the outgoing light.
The dependence of the radiative losses and of the far-field distribution
on the position and orientation of the defect are calculated.
The angular dependence of the far field is given by a lorentzian
with a width that has a sharp minimum for a certain optimal orientation
of the defect line. For this orientation the
whispering-gallery mode of a circular resonator is scattered by
the extended defect in the direction normal to the disk boundary.
\end{abstract}

\section{Introduction}

The idea to use a microdisk geometry as an alternative to the
Fabry-Perot cavity in a resonator design for a semiconductor laser was
introduced a decade ago\cite{mccall92}. The advantage of this geometry
is that the losses for the whispering-gallery modes of a circular
resonator are governed by evanescent leakage and, thus, can be very
low.  Namely, for a mode with a maximal angular momentum $M=nk_0R$,
where the $n$ is the effective refraction index,
 $R$ is the resonator radius, and $k_0$ is the wave number of the
 radiation, the quality factor, $Q$, with exponential accuracy
is given by
\begin{equation}
\label{Q1}
\ln Q =2k_0R\left[n\ln \left( n+\sqrt{n^2-1} ~\right) - \sqrt{n^2-1} ~
\right].
\end{equation}
The value of the
effective refraction index
is determined by the disk thickness
and the indexes of an active and surrounding passive layers.
In the pioneering paper Ref.~\onlinecite{mccall92} the effective index
was $n\approx 2$,
 while $k_0R $ for the smallest microdisk was $\approx 6$. Then
Eq.~(\ref{Q1}) yields $Q \approx 5 \cdot 10^5$.
 Experimentally  measured
values of $Q$ are much smaller, $Q\sim 150$, Ref.~\onlinecite{levi93}. The
discrepancy is partially due to a prefactor neglected in
Eq.~(\ref{Q1}), but primarily due to the absorption in
the active layer\cite{mccall92,slusher93}.
 With such a high
$Q$-value the lasing threshold for a microdisk resonator is very low.
For the same reason the output power is also low, which is not
desirable.  Another serious drawback of the microdisk geometry is that
the angular dependence of the output intensity is $I(\psi)\propto
\cos^2(M\psi)$, whereas applications require a directed emission.  In
order to remedy these drawbacks two proposals were put forward

\noindent ({\em i}) to extract the light out of the resonator by using
two parallel
disks\cite{chu94}. The first disk with high $Q$ contains a
multiple quantum
well structure in which the light is generated. The second passive
disk coupled to the laser contains an opening serving as
a leakage source. The shape of the opening determines the
directionality of the output light.

\noindent ({\em ii}) to couple the light out by introducing either an
identation in the form of the ``tip of the egg''\cite{levi93} or
corrugation\cite{li97} on the {\em circumference} of the disk.

A radical solution for increasing the output, and, to a certain
extent, directionality, by deforming the shape of the
disk\cite{gmachl98} seem to devaluate the attempts to extract light
from a perfectly circular microdisk. This solution
relied on the fact that deformation causes a
qualitative change in the light-ray dynamics, so that
the whispering-gallery trajectory of a ray becomes unstable.
As a result, the ray eventually impinges on the boundary at an
angle smaller than the critical angle, $\sin^{-1}(1/n)$.
This leads to a refractive escape.
The improvement of the directionality of the output light from a
wave-chaotic resonator was studied theoretically in a great
detail\cite{narimanov99,narimanov00}.
 The results of calculations for both ``bouncing ball'' and
``bow-tie'' modes and $nk_0R \approx 100$ can be roughly
summarized as follows. In each $90^{\circ }$-quadrant the
output light is concentrated within total angular interval
of about $60 ^{\circ }$ with a strong peak of  a width
$\sim 30 ^{\circ }$ and a large number of narrow
satellites\cite{narimanov99}.

In the present paper we suggest an alternative approach for
improving  both the directionality and the
output efficiency of a {\em circular} microdisk.
This  improvement can be achieved by
introducing a properly oriented {\em linear} defect
{\em away from the circumference}.
Proposed geometry is illustrated in Fig. 1. The reason why the
linear defect causes  directional emission from a microdisk is the
following.
The field of a  whispering-gallery mode ``tunnels'' {\em towards}
the defect line, which then assumes a role of the secondary source.
Since the source is extended, it emits a secondary light  beam which is
weakly divergent.
The divergence is minimal when this secondary light beam
is emitted in the radial direction, i.e. in the direction
normal  to the disk boundary.
It is convenient
to characterize
the position and orientation of the defect by two
parameters, namely
$r_0 \gg k_0^{-1}$ - radial distance from the edge to the circumference,
 and $d$ - the minimal distance from the defect line to the disk center.
 As it will be shown below, the optimal orientation of the defect,
for which the direction of the secondary beam is radial,
is determined by the condition $d=(R-r_0)/\sqrt{2}$. Under this
condition the directionality of the output light is maximal.
 Below we will demonstrate that, with exponential accuracy,
the radiative losses caused
by the defect are given by
\begin{equation}
\label{Q2}
\ln Q = \frac{2^{5/2}}{3} \left( \frac{r_0}{R} \right)^{3/2}  (nk_0 R) .
\label{q}
\end{equation}
These losses dominate over the evanescent losses Eq. (\ref{Q1}) if $r_0
\ll R$.
The angular dependence of the defect-induced emission is a lorentzian,
which under the optimal condition $d=(R-r_0)/\sqrt{2}$, has the form
\begin{equation}
\label{i}
I(\psi )=\frac{1}{\left( \psi - \mbox{\Large $\frac{\pi}{4}$} \right)^2
+ 2 n^2 \left(
\mbox{\Large $\frac{r_0}{R}$}
  \right)},
\label{L}
\end{equation}
with the width which is also governed by the ratio $r_0/R$. Note, that
although Eqs. (\ref{Q2}), (\ref{i}) apply only
for $k_0r_0 \gg1$, this ratio can still
be quite small as long as  $k_0R$ is large.

The paper is organized as follows. In Sec. 2 we
derive Eqs. (\ref{Q2}), (\ref{i}) within the scalar diffraction
theory.
In Sec. 3 we discuss the limits of applicability of the theory
and provide numerical estimates.

\section{Angular Dependence of the Output Light}

Neglecting the difference between TE and TM polarizations,
the field of a whispering-gallery mode in a microdisk represents
a solution of the two-dimensional Helmholtz equation
\begin{equation}
{\cal E}_M (\rho , \phi ) \propto \cos \left( M\phi \right)
J_M (nk_0 \rho),
\label{e1}
\end{equation}
where $\rho $ and $\phi $ are polar coordinates,
$M$ is the angular momentum,
and $J_M$ is the Bessel function.
We assume that $M$ is close to the maximal value $nk_0 R$.
Then the field
Eq.~(\ref{e1})  is localized at the boundary $\rho = R$
within a narrow ring of a width
$\delta \rho \sim R/(nk_0 R)^{2/3} \ll R$.
At smaller $\rho $ the field falls off towards the center of the disk as
\begin{equation}
{\cal E}_M   \propto \cos \left( M\phi \right)
         \exp \left[ -\frac{2^{3/2}}{3M^{1/2}}  \left( nk_0 r
              \right)^{3/2} \right],
\label{f1}
\end{equation}
where $r = R - \rho$ is the distance from the boundary.

Within the  scalar diffraction theory the emitted field caused by the
presence of a defect is determined by the Fresnel-Kirchhoff diffraction
integral
in which the source is the field
${\cal E}_M (\rho , \phi )$ taken at $\rho = \rho (\phi )$, where
$\rho (\phi )$ describes the defect profile. In the case of a linear
defect (Fig.~1) we have $\rho (\phi )= d/\cos \phi $.
It is convenient to introduce instead of $\rho $ a variable $x$ which
is the distance along the defect (Fig.~1). The relation between
$\rho $ and $x$ is the following
\begin{equation}
\rho =  \left[ d^{~\!2} + \left( \sqrt{(R-r_0)^2 - d^2 } -x  \right)^2
 \right]^{1/2} =
   R - r_0 - x \frac{ \sqrt{ (R-r_0)^2 -d^2}}{R-r_0}  .
\label{rr1}
\end{equation}
In the second equality we have used the fact that $x \ll R$.
Substituting Eq.~(\ref{rr1}) into Eq.~(\ref{f1}) we obtain
\begin{equation}
{\cal E}_M (x)  \propto  \exp\left[
    -\frac{2^{3/2}}{3}\left(\frac{r_0}{R} \right)^{3/2}
   nk_0 R \right] \exp (-a x ) \cos [ M \phi (x) ] ,
\label{fm1}
\end{equation}
where
\begin{equation}
a = 2^{1/2} nk_0 R \left(\frac{r_0 }{R^3} \right)^{1/2}
      \frac{\sqrt{(R-r_0)^2 - d^2}}{R-r_0} .
\label{a1}
\end{equation}
In Eq.~(\ref{fm1}) we assumed that $x \ll r_0$. Indeed,
the relevant values of $x$ are $\sim a^{-1}$. Then the
condition $x \ll r_0$ can be rewritten as $r_0 a \sim
(nk_0 R) (r_0/R)^{3/2} \gg 1$. We see that this condition
is equivalent to the requirement that the asymptotic
Eq.~(\ref{f1}) is valid at $r = r_0$.
The first $x$-independent factor in Eq.~(\ref{fm1}) determines the
dependence of the output field on the defect position, $r_0$. The
expression Eq.~(\ref{q}) immediately follows from this dependence.

The form of the function $\phi (x)$ in Eq.~(\ref{fm1})
can be easily established from Fig.~1
\begin{equation}
 \tan \phi =  \frac{\sqrt{(R-r_0)^2-d^2} -x}{d}  .
\label{phi}
\end{equation}
Now we are in position to write the expression for
the intensity of the outgoing light in the direction
$\psi $. It is given by the following integral along
the defect
\begin{equation}
I (\psi )  \propto \left| \int _0 ^{\infty } ~dx~ e^{-a x}
    \cos \left[ M \phi (x) \right]
 \int _{-\pi}^{\pi } d\varphi ~ \exp \Bigl[ i nk_0 b(x,\varphi ) -
 i k_0 R \cos(\psi - \varphi)   \Bigr]
  \right|^2.
\label{I1}
\end{equation}
The internal integral over $\varphi $ is
a standard
Fresnel-Kirchhoff integral. Parameter $b$ in the exponent
is the distance from the source on the defect to the exit
point (Fig.~1)
\begin{equation}
b^2(x, \varphi ) = R^2 + d^{~\!\!2} \!\cos^2 \left[ \phi (x) \right] -
  2 ~\!\! R ~ \! d \cos \left[ \phi (x) \right] \cos \left[ \varphi -
\phi (x)
    \right].
\end{equation}
It is convenient to express the distance $b$ directly through
$x$ and $\varphi $, which can be done using Eq.~(\ref{phi})
\begin{equation}
b^2(x, \varphi ) = R^2 +(R-r_0)^2 -2R \left( d \cos \varphi +
\sqrt{(R-r_0)^2 -d^2}~\sin \varphi
   \right)
+2 x \left( R \sin \varphi - \sqrt{(R-r_0)^2 -d^2} \right) + x^2.
\end{equation}
Recall now, that the values of $x$ in the integral Eq.~(\ref{I1})
are small
 $x \sim a^{-1} \ll r_0$.
It can  also be seen from Fig.~1 that the outgoing ray is normal
to the boundary when $\cos \varphi = d/(R-r_0)$. This suggests
that the difference
\begin{equation}
\delta  = \varphi - \cos^{-1} \left( \frac{d}{R-r_0 } \right)
\end{equation}
is a small parameter. In other words, the major contribution
to the Fresnel-Kirchhoff integral comes from small $\delta \ll 1$.

The integrand in Eq.~(\ref{I1}) is a rapidly oscillating
function.
This allows to expand the phase of the oscillations
\begin{equation}
\Phi (x, \varphi ) = M\phi (x) +k_0 \left[ n b(x, \varphi )-
    R \cos(\psi -\varphi )\right]
\end{equation}
in terms of $x$ and $\delta $
\begin{equation}
\Phi (x, \varphi ) = A_x x  +
 A_{xx} x^2 + 2 A_{x  \delta } ~ \! x \delta +  A_{\delta \delta } ~\!
\delta ^2.
\label{AA}
\end{equation}
As it was already stated in the Introduction, the maximal
directionality of the outgoing light is achieved for the position
of the defect $d = (R-r_0)/\sqrt{2}$. To demonstrate this, we
introduce a dimensionless deviation from the optimal defect
position
\begin{equation}
 \Delta (d)  =  \frac{d}{R} - \frac{R-r_0}{\sqrt{2}R } .
\label{delta}
\end{equation}
We will see that the width of the function $I(\psi )$ increases
dramatically with $\Delta $.
Rather involved but straightforward calculations yield the following
expressions for the coefficients in the expansion Eq.~(\ref{AA})
\begin{eqnarray}
& & A_{x \delta } = \frac{ n k_0 R}{  2 r_0}
\left(1- \frac{4 \Delta ^2}{1- 4 \Delta ^2}\right) \approx
 \frac{ n k_0 R}{  2 r_0}
\left(1- 4 \Delta ^2\right) ,  \label{A1} \\
& & A_{\delta \delta } =  -  k_0 R   \left[ 1 - \frac{nR}{r_0}
\left( 1- \frac{4 \Delta ^2}{1- 4 \Delta ^2}\right)  \right] \approx
  -  k_0 R   \left[ 1 - \frac{nR}{r_0}
\left( 1- 4 \Delta ^2 \right)  \right] ,\\
& &  A_{xx} = \frac{ n k_0}{4 r_0} ~~~~~~, ~~~~
  A_x = a \frac{\psi - \pi/4}{\delta  _{\psi }}  , \label{A2}
\end{eqnarray}
where the parameter $\delta _{\psi } $ in the expression for
$A_x $ is defined as
\begin{equation}
\delta  _{\psi } = n \left( \frac{2r_0}{R} \right) ^{1/2} \!
\left[ \frac{1-4 \Delta  ^2}{1-4n^2 \Delta ^2} \right] ^{1/2}
\approx  n \left( \frac{2r_0}{R} \right) ^{1/2} \Bigl[ 1+2(n^2-1)
   \Delta  ^2 \Bigr].
\end{equation}
With the use of the expansion Eq.~(\ref{AA}), the Fresnel-Kirchhoff
integral can be easily evaluated yielding
\begin{equation}
I (\psi )   \propto  \left| \int _0 ^{\infty } ~dx~
\exp \left[ -x \left(a- i A_x \right)
    +i x^2 \left( A_{xx} - \frac{A_{x \delta }^2 }{A_{\delta \delta }}
\right) \right]
  \right|^2 .
\label{I3}
\end{equation}
The remaining integral over $x$ is of the Fresnel-type. However, it
 cannot
be reduced to the special functions $Ci(u)$ and $Si(u)$, which describe
the diffraction from a semi-infinite plane\cite{landau}. This is
because the linear term in the exponent contains a  contribution $-ax$
which is  {\em real}. For this reason, it is convenient to introduce
a new variable $z= ax$ in the integral (\ref{I3}). Upon substituting
the coefficients (\ref{A1})-(\ref{A2}) into Eq.~(\ref{I3}) we arrive
at the final result
\begin{equation}
I (\psi )  \propto \left| \int _0 ^{\infty } ~dz~
\exp \left[ -z \left(1- i  \frac{\psi -\pi/4}{\delta _{\psi }}\right)
    +i z^2\frac{n+1}{4n^2 k_0 r_0 } F(\Delta ) \right]
  \right|^2 ,
\label{I2}
\end{equation}
where the function $F(\Delta )$ is defined as
\begin{equation}
 F( \Delta   ) = 1 +  \frac{8 n R \Delta ^2}{(n+1)r_0 }.
\end{equation}
As in  Eqs.(\ref{A1})-(\ref{A2}), we kept only the leading $\Delta ^2$
term in the definition of $F$.
Now we can substantiate the statement that the optimal directionality
of the emission is achieved at $\Delta =0$. Indeed, $z^2$-term
in the exponent of Eq.~(\ref{I2}) leads to the broadening and
oscillations
of the angular dependence, $I(\psi )$. At small $\Delta $ we have
$F\approx 1$; then the $z^2$-term is multiplied by a small factor
$\sim(k_0 r_0)^{-1} \ll 1$ and, thus, can be neglected.
Then we immediately recover the lorentzian Eq.~(\ref{L}). On the
other hand, for a general position of the defect
we have $\Delta \sim 1$,
and $F \sim R/r_0 $. Then the $z^2$-term acquires a much larger
coefficient $2R/(nk_0r_0^2)$, resulting in the loss of the
directionality
of the output light. This is illustrated in Fig.~2. It is seen that
significant broadening and sideback oscillations set in already
at  small values of $\Delta $. In particular, for $\Delta = 0.3$
the broadening is 60 percent.

\section{Conclusion}

 Let us first discuss the validity of the assumptions
used in the above calculation

\noindent
{\em (a)} $I(\psi )$ was calculated within the scalar
diffraction theory using Fresnel-Kirhhof approach. Note, that
for a circular geometry, $I(\psi )$ can be calculated exactly by solving
the scalar wave equation and treating defect as a perturbation.
Then the
expression for $I(\psi )$ is given by
a sum over angular momenta of the leaking modes.
Fresnel diffraction corresponds to
replacing  this sum by an integral.
The accuracy of
such a replacement is determined by the next
term in the Poisson expansion, which contains an exponential
factor $\exp \left[ - 2^{3/2} \pi n k_0 (r_0 R)^{1/2} \right]$.
Thus, the condition of validity of the Fresnel-ËKirhgof approach
is
$r_0 \gg 1/(k_0^2 R)$, which is not restrictive at all.

\noindent
{\em (b)}
According to Eq.~(\ref{L}), the full width at half maximum
(FWHM) is equal to $2\delta _{\psi } = 2 n (2r_0 /R)^{1/2} $.
This equation was derived under the assumption that the
defect is located far enough from the circumference of the
disk, {\em i.e.} $r_0 \gg \delta \rho \sim R/(nk_0 R)^{2/3}$.
It is possible to derive a more general
expression for $I(\psi )$,
that is valid
for $r_0 \sim  \delta \rho $, when the
asymptotics Eq.~(\ref{f1}) is not yet applicable. Derivation
is based on the integral representation of the Bessel
function and yields
\begin{equation}
I(\psi ) \propto  \frac{2}{( \pi  \gamma )^{1/2}}
   \int _0 ^{\infty } ds \frac{e^{- \gamma s^2 }  }{
   (1+  \mbox{\large $s$})^2 +  \left( \frac{
\mbox{\large $\psi$} - \mbox{\large $\pi /4$ }  }{\mbox{
\large $\delta _{\psi }$}} \right)^2  },
\end{equation}
where the  parameter $\gamma $ is defined as
\begin{equation}
\gamma = 2^{1/2} nk_0 R  \left( \frac{r_0}{R} \right)^{3/2}.
\label{gamma}
\end{equation}
It is seen that the  condition $r_0 \gg \delta \rho $ corresponds to
$\gamma \gg 1$. Then we immediately recover the Lorentzian
Eq.~(\ref{L}). At moderate $\gamma $, the FWHM
is given by $2 C(\gamma ) \delta _{\psi }$, where the function
$C(\gamma )$ is plotted
in Fig.~2, inset.  It is seen that
within the wide interval $1 \lesssim \gamma \lesssim 10 $
the broadening factor $C(\gamma )$ changes very slowly. Then
the FWHM can be expressed in terms of $\gamma $ as
$2^{4/3} n C(\gamma ) \left( \gamma / nk_0 R \right)^{1/3}$,
which is also a slow function of $\gamma $. Choosing for
concreteness $\gamma =1$, we find for FWHM the
expression $3.35 n/(n k_0 R)^{1/3}$.

We now turn to the numerical estimates.
Three types of microdisk semiconductor lasers
have been described
in the literature so far.
The lasers for wavelengths $\lambda \approx 1.5 \mu m$ have $M$-values
reported\cite{mccall92,levi93,slusher93,chu94,mohideen94,chu95,zhang95,lee98,baba00,mohideen'94}
 are rather low
($10\lesssim M \lesssim 70$) and $n\approx 2.5$. For this $n$ and
maximal $M=70$ the FWHM is $116^{\circ }$.
 Lasers for $\lambda \approx 0.8 \mu m $\cite{backes,ahn99,park01}
have
$n \approx 3.1$ and  also rather small $M$
($30 \lesssim M \lesssim  300$). With maximal $M=300$ we get
$89^{\circ }$ for  FWHM.
Nitride-based lasers operating at
$\lambda \approx 0.4 \mu m$\cite{mair,schang,zeng}
 have much higher $M$-values
($200 \lesssim M \lesssim  600$) and $n \approx 2.8$.
This yields FWHM of $64^{\circ }$.
Microdisk lasers based on non-crystalline materials
(polymer\cite{polson00} and dye solution\cite{moon}) have also been
reported.
For this materials $n\approx 1.8$ is smaller and the values
of $M$ ($930$ in \cite{polson00} and 3000 in \cite{moon}) are high.
Both factors tend to narrow $I(\psi )$. Namely, for $M=1000$ the  FWHM
of $34^{\circ }$  can be achieved.

Let us discuss the physical meaning of the optimal condition,
$d= (R-r_0)/\sqrt{2}$. As it is seen from Eq.~(\ref{fm1}),
the phase of the whispering-gallery mode changes along the defect.
As the defect plays a role of a source of the outgoing light,
this change, $\phi (x)$, is equivalent to the rotation of the
line of the constant phase by an angle $\sin^{-1} [ d/(R-r_0)] $.
Then, under the optimal condition, the line of the constant phase
is perpendicular to the radial line drawn through the edge of
the defect (Fig.~1). In other words, under the optimal condition,
the defect can be replaced by a constant phase line at distance
$r_0$ from the circumference
that is {\em parallel} to the circumference.
Clearly, the angular width of the far field emitted by this line
is minimal for this parallel orientation.

Note in conclusion, that in \cite{gmachl98} the improvement of
the output characteristics of microdisk laser,
achieved by introducing the deformation, is due to the
fact that when the disk is deformed, the light rays
are unable to stay within a whispering-gallery trajectory, and
experience refractive escape in course of the
chaotic motion\cite{stone97}.
In the present paper we considered a perfectly circular
microdisk with a defect. A point-like
defect at some distance away from the boundary would be unable
to couple out all the whispering-gallery modes,
 since it will not be able to affect the modes
having a node at the  defect position. Our main message here is that
no whispering-gallery mode can evade an {\em  extended}
defect and will be directed out of the resonator as a result of
scattering by this defect.

\begin{figure}
\centerline{
\epsfxsize=3.4in
\epsfbox{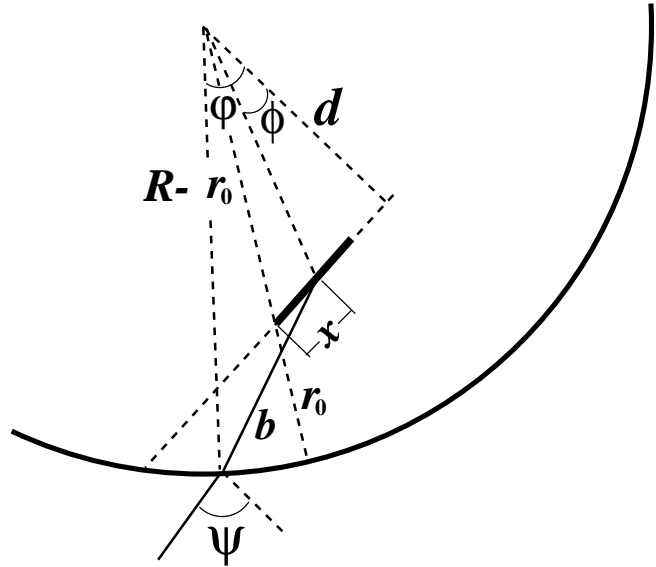}
\protect\vspace*{0.1in}}
\protect\caption[fig1]
{\sloppy{ Shematic illustration of a circular microdisk
of a radius $R$ with a linear defect. The defect position
is characterized by $r_0$ - the distance from edge to
the disk circumference along the radius; the defect
orientation is fixed by the minimal distance from the
defect line to the disk center. The direction of the
outgoing light is characterized by the angle $\psi $.
}}
\label{figone}
\end{figure}

\begin{figure}
\centerline{
\epsfxsize=4.2in
\epsfbox{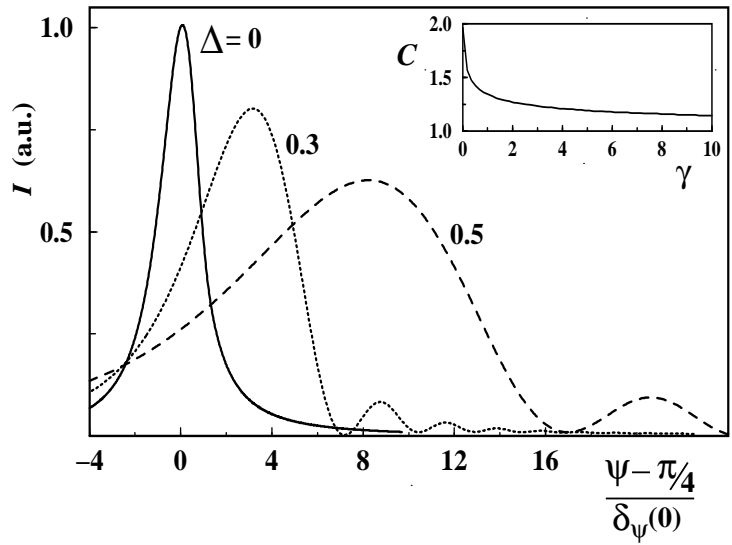}
\protect\vspace*{0.1in}}
\protect\caption[fig2]
{\sloppy{ Angular distribution of the far-field
emission intensity is plotted for different deviations
$\Delta $ (Eq.~(\ref{delta})) from the optimal condition
$d = (R-r_0)/\sqrt{2}$. Inset: dimensionless broadening
factor $C$ is plotted versus the dimensionless parameter
$\gamma $, defined by Eq.~(\ref{gamma}).
}}
\label{figtwo}
\end{figure}


\begin{references}

\bibitem{mccall92}S. L. McCall, A. F. J. Levi, R. E. Slusher,
S. J. Pearton,
and R. A. Logan, Appl. Phys. Lett. {\bf 60}, 289 (1992).


\bibitem{levi93}A. F. J. Levi, R. E. Slusher, S. L. McCall,
S. J. Pearton, and R. A. Logan, Appl. Phys. Lett. {\bf 62}, 561 (1993).

\bibitem{slusher93}R. E. Slusher, A. F. Levi, U. Mohideen, S. L. McCall,
S. J. Pearton, and R. A. Logan, Appl. Phys. Lett. {\bf 63}, 1310 (1993).



\bibitem{chu94}D. Y. Chu, M. K. Chin, W. G. Bi, H. Q. Hou, C. W. Tu,
and S. T. Ho, Appl. Phys. Lett. {\bf 65}, 3167 (1994).

\bibitem{li97}B.-J. Li and P.-L. Liu, IEEE J. Quant. Electron. {\bf 33},
791 (1997).


\bibitem{gmachl98}C. Gmachl, F. Capasso, E. E. Narimanov,
J. U. N\"{o}ckel,
A. D. Stone, J. Faist, D. L. Sivco, and A. Y. Cho, Science {\bf 280},
1493 (1998).

\bibitem{narimanov99}E. E. Narimanov, G. Hackenbroich, P. Jacquod, and
  A. D. Stone, Phys. Rev. Lett. {\bf 83}, 4991 (1999).

\bibitem{narimanov00}O. A. Starykh, P. R. J. Jacquod, E. E. Narimanov,
and A. D. Stone, Phys. Rev. E {\bf 62}, 2078 (2000).



\bibitem{landau} L. D. Landau and E. M. Lifschitz,
{\em The Classical Theory of Fields}
  4th edition, Pergamon Press, Oxford, (1975).

\bibitem{mohideen94}U. Mohideen, R. E. Slusher, F. Jahnke,
and S. W. Koch, Phys. Rev. Lett. {\bf 73}, 1785 (1994).

\bibitem{chu95}D. Y. Chu, S. T. Ho, X. Z. Wang, B. W. Wessels, W. G. Bi,
C. W. Tu, R. P. Espindola, and S. L. Wu, Appl. Phys. Lett. {\bf 66},
2843 (1995).
\bibitem{zhang95}J. P. Zhang, D. Y. Chu, S. L. Wu, S. T. Ho, W. G. Bi,
C. W. Tu, and R. C. Tibero, Phys. Rev. Lett. {\bf 75}, 2678 (1995).

\bibitem{lee98}T.-D. Lee, P.-H. Cheng, J.-S. Pan, R.-S. Tsai, Y. Lai,
and K. Tai, Appl. Phys. Lett. {\bf 72}, 2223 (1998).

\bibitem{baba00}T. Baba, H. Yamada, and A. Sakai, Appl. Phys. Lett.
{\bf 77}, 1584 (2000).

\bibitem{mohideen'94}U. Mohideen, W. S. Hobson, S. J. Pearton, F. Ren,
and R. E. Slusher, Appl. Phys. Lett. {\bf 64}, 1911 (1994).

\bibitem{backes} S. A. Backes, J. R. A. Cleaver, A. P. Heberle,
J. J. Baumberg, and K. K\"{o}ler, Appl. Phys. Lett. {\bf 74},
176 (1999).

\bibitem{ahn99}J. C. Ahn, K. S. Kwak, B. H. Park, H. Y. Kang, J. Y. Kim,
and O'Dae Kwon, Phys. Rev. Lett. {\bf 82}, 536 (1999).

\bibitem{park01} B. H. Park, J. C. Ahn, J. Bae, J. Y. Kim, M. S. Kim,
S. D. Baek, and O'Dae Kwon, Appl. Phys. Lett. {\bf 79}, 1593 (2001).

\bibitem{mair}R. A. Mair, K. C. Zeng, J. Y. Lin, H. X. Jiang, B. Zhang,
L. Dai, A. Botchkarev, W. Kim, H. Morkoc, and M. A. Khan, Appl. Phys.
Lett. {\bf 72}, 1530 (1998).

\bibitem{schang}S. Chang, N. B. Rex, R. K. Chang, G. Chong,
and L. J. Guido, Appl. Phys. Lett. {\bf 75}, 166 (1999).

\bibitem{zeng}K. S. Zeng, L. Dai, J. Y. Lin, and H. X. Jiang, Appl.
Phys. Lett. {\bf 75}, 2563 (1999).

\bibitem{polson00}R. C. Polson, G.  Levina, and Z. V. Vardeny,
 Appl. Phys. Lett. {\bf 76}, 3858 (2000).

\bibitem{moon}H.-J. Moon, Y.-T. Chough, J. B. Kim, K. An, J. Yi, and
J. Lee, Appl. Phys. Lett. {\bf 76}, 3679 (2000).

\bibitem{stone97} J. U. N\"{o}ckel and A. D. Stone, Nature {\bf 285},
 45 (1997), and references therein.





\end{references}
\end{document}